\documentclass[preprint,12pt]{aastex}
\usepackage{epsfig}
\begin{document}
\title{
Short-duration lensing events: II. Expectations and Protocols
}
\author{Rosanne Di\thinspace~Stefano}
\affil{Harvard-Smithsonian Center for Astrophysics, 60
Garden Street, Cambridge, MA 02138}


\begin{abstract}
Ongoing microlensing observations by OGLE and MOA
regularly identify and conduct high-cadence sampling of
lensing events with Einstein diameter crossing time,
$\tau_E,$ of $16$~or fewer days.
Events with estimated values
of $\tau_E$ of one to two days have been detected. 
Short duration events tend to be generated by low-mass lenses
or by lenses with high transverse velocities. We compute the expected rates,
demonstrate the expected ranges of parameters for lenses of different mass,
and develop a protocol for
observing and modeling short-duration events.
Relatively minor additions to the procedures presently used
will  
increase the rate of planet discovery, and also discover or
place limits on the population of high-speed dim stars and stellar  
remnants in the vicinity of the Sun.
\keywords{planetary systems; stars: low mass, brown dwarfs; stars:neutron; white dwarfs; solar neighborhood; Galaxy: halo}
\end{abstract} 

\section{Introduction} 
\def\ev{event}  

The work described in this paper 
is inspired by the success achieved by the OGLE
 (Udalski 2003) and MOA (Bond et al.\, 2001)
teams in discovering and monitoring events of short duration.
Short-duration events now constitute a significant fraction of
all event candidates.  
{For example, approximately $9\%$ of $654$ recent events
listed by the OGLE team on its
{\it Early Warning Site} 
 {\it(EWS;  http://ogle.astrouw.edu.pl/ogle3/ews/ews.html)} have
Einstein-diameter crossing times shorter than $8$~days. Twenty-seven
percent of these short events have $\tau_E < 4$~days, and a handful
have $\tau_E < 2$~days.}
While not all of the candidate events of short duration are
likely to be lensing \ev s, many of the light curves appear to be well fit
by lensing models. 
The prospects for increasing the rate of discovery of short-duration 
events are good. Ongoing programs are continuing to implement improvements,
while new projects are starting or being planned.

The value of $\tau_E$ is an estimate of the time during which the
gravitational magnification would exceed $1.34$ for close 
approaches between the source and lens. 
Current observing
programs are sensitive enough to detect ongoing events when the
magnification is just a few percent, increasing the effective event
duration by a factor that can be as large as $3.5$.\footnote{For heavily blended events, or events in which the peak magnification is low, the enhancement
factor will be smaller (Di\thinspace Stefano \& Esin 1995).}
 These programs are therefore able to call alerts
on ongoing events with small values of $\tau_E$.
 Just as alerts on events deemed likely to
produce caustic crossings inspire world-wide ``follow-up''
with more frequent observations (see Griest \& Safizadeh 1998), 
alerts on short-duration events
can be accorded high priority. The goal is to collect enough data 
to permit detailed model fits.  

The scientific advances made possible by studying short events
are significant. 
The systematic study of short-duration events will yield information
about brown dwarfs, planets orbiting stars, as well as 
free-floating planets\footnote{Any planet-mass 
object not bound to a star will be referred to as a 
{\it free-floating planet} in the rest of the text. 
Some such low-mass objects may have been
formed in and ejected from planetary systems, and others may have been
formed in isolation.} 
and hypervelocity objects. 
In the companion paper we explored the 
benefits of selecting for intensive study events of short
duration (Di\thinspace Stefano 2009). Here we
 develop strategies that can help monitoring 
programs to discover planets, brown dwarfs, and hypervelocity objects.

In \S 2 we estimate the
rate of such events, relative to the rate of all detected events in
both present and future monitoring programs. We demonstrate that the
rate is high enough that existing data contain evidence of
 hundreds of interesting short-duration events. More important, every year
an additional $> 100$ short-duration events should be
detected. These can be followed 
in real time, maximizing the science return from each. In \S 3 we
consider the relationship between event characteristics and the location
of the lens, and give some explicit examples of model tests that are
possible under
ideal circumstances.
These brief sketches integrate the measurements 
discussed in the companion paper.   
Section~4 reviews the prospects for 
optimizing, in the immediate future, what we can learn about
planet lenses and the other intriguing masses that
can generate events of short duration.

In the appendix we point out that short-duration events 
events caused by wide-orbit planets, free-floating planets, and
hypervelocity objects
will be augmented by short-duration
events expected when the lens is a planetary system with close-in
planets. The orbital separations in some of these additional cases
put the planets in the habitable zone (Di~Stefano \& Night 2008).

\section{The Rate of Short-Duration Events} 

\subsection{Estimates}

The populations 
producing short-duration events are comprised of objects that are
not yet well-studied. Predictions of the rates at which they should produce
events therefore have large uncertainties. We have used a simple
and straightforward approach which allows the effects of 
each assumption to
be traced, so that adjustments can be easily made if necessary.   

We assumed that the
majority of lensing events presently detected are generated by stars,
and that the majority of stellar-lens events are generated by M~dwarfs.
If the rate at which M-dwarf events are detected is ${\cal R},$ then the 
rate of detected events caused by other types of lenses can be
obtained by scaling according to lens mass,
transverse velocity, and spatial density. 
This was carried out in Di~Stefano 2008b for brown dwarfs and 
stellar remnants.   
Those calculations produced the relative rates of events 
shown in column 1 of Table 1, for 
brown dwarfs, white dwarfs, neutron stars, and black holes. 

Main sequence stars more massive than $M$ dwarfs also cause lensing events.
To compute the rate at which they generate events, relative
to the rate at which M~dwarfs generate events, 
we simulated a population of stars using the Miller-Scalo mass function.  
By using the fact that 
the rate at which each star produces lensing events scales as the square root of its mass,
we found that stars with masses above $0.5\, M_\odot$ produce approximately as many events 
as stars of lower mass, even though the mass function 
declines with increasing mass.   
 Because, however,  
more massive stars are more luminous than M~dwarfs, 
events generated by them can be more difficult
to detect, even if image differencing is employed. We therefore assume that, while
events by both nearby and more distant M~dwarfs contribute to the events
discovered by the OGLE and MOA teams, only more distant stars of larger mass contribute. 
We therefore took the total rate of detectable
lensing by stars to be $1.5\, {\cal R},$
with stars more massive than M~dwarfs contributing $1/3$ of the total.

In addition, planets in wide orbits produce events. 
The average rate per star is 
$\sqrt{M_\ast}\, \sum_i \sqrt{q_i},$ where the sum is over
the number of wide-orbit planets, and 
$q_i=M_{planet,i}/M_\ast$. 
Here we will assume that the rate of planet-lens events is
$15\%$ the rate of M~dwarf events.
The rate at which the outer planets of our solar system generate
events, compared to the rate of events due to a star of $0.25\, M_\odot$
is roughly $0.12.$ Many stars appear to have planets more massive than Jupiter,
however, and some of these appear to be in very 
wide orbits\footnote{http://exoplanet.eu/}. 
Free floating planets are likely to add a significant contribution.
Although the value of $15\%$ is uncertain,
the true rate seems likely to lie 
within a factor of $2$ of it. 

As the companion paper demonstrates, planet-lens events are likely to be the  
shortest events. They are therefore more likely to be missed
by present-day monitoring teams. 
 In Table~1 we assume that the efficiency for
detecting events by planet-mass lenses is only half that  
for longer-lasting events. As we will discuss in \S 4, however, some future
programs will improve the efficiency of detecting short-duration
events.  
In Table~1 we 
therefore include a column for ``future'' monitoring programs
which achieve equally good efficiencies for short and long events.

Among high-velocity objects, we have included computations 
only for neutron stars, because we can make rough but reasonable estimates
based on the inferred galactic population of neutron stars. 
 Other high-velocity objects 
can be expected to supplement these numbers; 
these include dim halo stars and remnants of runaway and hypervelocity stars. 
\begin{table*}
\centering{
\caption{Numbers of Short-Duration Events ($\tau_E < 16$~days)}
\smallskip
\begin{tabular}{|c||c|c|c|c|c|c|}
\hline
              &          &Number        &Number   &Number      &Number\\
              &Relative  &of events     &of short events&of events&of short events\\
 Lens         &Event     &per 1000      &per 1000 &per 1000    &per 1000\\
 Type         &Rate      &detected      &detected &detected    &detected\\
              &   &{\sl Present}$^{(1)}$&{\sl Present}&{\sl Future}$^{(2)}$&{\sl Future}\\
\hline
M dwarfs      &  $1.0$   &  $481$       &$5^{(3)}$ & $465$      &  $5$  \\ 
Other stars   &  $0.5$   &  $241$       &   $0$   & $233$      &  $0$  \\ 
Brown dwarfs  &  $0.19$  &  $92 $       &   $69$  & $88$       &  $66$ \\ 
White dwarfs  &  $0.17$  &  $82 $       &   $0$   & $79$       &  $0$  \\ 
Neutron stars &  $0.13$  &  $63 $       &   $13$  & $60$       &  $12$ \\ 
Black holes   &  $0.01$  &  $5  $       &   $0$   & $5$        &  $0$  \\ 
Wide-orbit planets&$0.15^{(4)}$& $36$    &   $36$  & $70$       &  $70$ \\  
\hline
\end{tabular}
}
\par
\medskip
\begin{minipage}{0.94\linewidth}
\footnotesize
\smallskip 
NOTES:  $^{(1)}$ {\sl Present} refers to the number of events 
generated by each type
of lens per $1000$ events detected by monitoring programs with the capability of
OGLE III.
$^{(2)}$ {\sl Future} refers to the number of events 
generated by each type
of lens per $1000$ events detected by monitoring programs of the future which
will be more sensitive to events with $\tau_E < 2$~days.
$^{(3)}$ We assume that roughly $1\%$ of all events generated by M~dwarfs
will have $\tau_E < 16$~days). These correspond to very fast transverse
velocities or M-dwarf lenses very close to us. More massive stars 
that are close to us may be too bright to generate events detectable
by the monitoring programs. 
$^{(4)}$ We assume that only half of these are long enough to be detected
by today's monitoring programs, but that programs of the future,
such as KMTNet (see \S 4.2) will be able to detect most planet-lens events.   
\end{minipage}
\end{table*}
\subsubsection{Implications}

The numbers of events predicted in Table~1, though approximate,  
provide guidance for what we can expect from existing and future 
data sets. 
Column 4 of Table 1 shows that we can expect $124$
short events among each $1000$ events (Column 3). We 
therefore expect that, at present, $\sim 12\%$ of all detected events
should have $\tau_E < 16$~days. 
This is roughly consistent with
the data on the OGLE team's web site, although we don't know the fraction
of the posted events actually associated with lensing.
Nevertheless, it is likely that existing data sets, which include 
more than $4000$ event candidates, contain evidence
for roughly $500$ short-duration events.
Seventy percent of these are likely 
to have durations in the range $8-16$~days.
Of these, 
$\sim 80\%$ were caused by brown dwarf lenses, and the remainder by
high-velocity stellar remnants. 
The majority of the shorter events were caused by planets. 

Although we do not have an opportunity to improve upon
 the sampling of events that have already occurred, we can nevertheless
conduct meaningful tests and identify the natures of at least some of
the lenses. 
We should check existing multiwavelength catalogs and data sets
at the position of each event well-sampled enough that we are reasonably
confident (a)~it corresponds to microlensing and (b)~that 
the short duration is not
due to blending. In at least some cases we should find evidence for
an object that is not the lensed source, but which could instead be 
part of the lens system.   In cases where there is a match, the possible 
contributions of additional observations, including some with HST,
 should be considered.
It would be surprising if a comprehensive analysis of already-discovered
short events does not reveal the presence of a set of interesting lenses.  
 
The existing data  sets are almost certainly minute in comparison
with future data sets, which will discover more events per year. 
Column 6 of Table 1 shows that we can expect $150$
short events among each $1000$ events (Column 5) detected by
monitoring programs sensitive to short events. 
The challenge ahead is, therefore, to
institute real-time recognition of short-duration events to
immediately identify those that are promising candidates for additional
observations, We discuss this further in \S 6. 

\section{Learning about the Lens}  
\subsection{Lens Location and Finite-Source Size}

For each type of lens defined by a given mass range, the selection of 
short-duration events corresponds to a selection of  
lenses in one of two
distance regimes. 
This is because
the equation for $D_L$ is quadratic, generally admitting
two solutions, $D_L^+$ and $D_L^-.$  
\begin{equation} 
D_L^{\pm} = \Bigg({{D_S}\over{2}}\Bigg)\, 
\Bigg[1\, \pm \, \sqrt{1-4\, 
\Big[{{\tau_E}\over{0.68\, {\rm d}}}\, {{v}\over{75\, {\rm {{km}\over{s}}}}}\Big]^2
\Big[{{8\, {\rm kpc}}\over{D_S}}\Big]\, \Big[{{M_\earth}\over{M}}\Big]}  
\Bigg]
\end{equation} 
For lenses that are brown dwarfs, planets, and neutron stars, 
$D_L$  is shown as a function of $v$ in Figures $1$ through
$3$, respectively. 
$D_L^-$ can be in the range of tens or hundreds of pc. When the lens system
is this  close to us, the probability of being able to detect it is large.  
As discussed in the companion paper, 
the degeneracy inherent in lensing
can therefore often be broken. 
In fact, for a given lens, there may
be several different ways of measuring the some key quantities, such as the 
mass of the planet-lens (see, e.g., \S 3.2). 

Because nearby lenses can be so well studied, we can use them to make 
predictions for the population of distant lenses, with $D_L=D_L^+.$
Nearby lenses producing short-duration events can be
identified as planets, brown dwarfs, or
stellar remnants. If we assume that stellar populations in the
dense source systems contain similar populations, we can predict the
distributions of values of $\tau_E$ and also of values of $\theta_E.$
($\theta_E$ is more likely to be measured for $D_L=D_L^+$). 
Comparisons between the predicted and observed distributions
will allow us to test models.
  
Note, in addition,  
that for a given lens mass and speed, the requirement that
$D_L$ be real, places an upper limit on the Einstein diameter
crossing time.
\begin{equation} 
\tau_E< 0.34\, {\rm d}\, \Bigg({{75\, {\rm {{km}\over{s}}}}\over{v}}\Bigg)\, 
\Bigg[{{M}\over{M_\earth}}\Bigg]^{{1}\over{2}} 
\Bigg[{{D_S}\over{8\, {\rm kpc}}}\Bigg]^{{1}\over{2}} 
\end{equation}
Thus, for the 
values of $v$ expected for planets, 
Earth-mass planets (dwarf planets) 
can produce only events shorter 
than about a day (an hour), 
while events generated by Jupiter-mass planets should have
$\tau_E$ less than roughly $18$~days.
In order for higher-mass lenses (such as brown dwarfs) to produce short
events, they must be very nearby or else very close to the
source 
(Figures 1 and 3). Although there are fewer lenses in these small volumes,
each of the nearby lenses has a relatively 
high probability of generating an event because values of both $\theta_E$ and
$\omega$ tend to be high (Di\thinspace Stefano 2008b).  

\subsection{Examples}
In many cases, 
particularly for mesolenses, 
it is possible to 
productively follow several different lines of evidence to learn more about
the lens system. In the companion paper we considered free-floating planets.
Below we consider examples of brown dwarfs, bound planets, and high-velocity
stars.

\subsubsection{Brown Dwarfs}

The first examples we consider are nearby brown dwarf lenses. 
It is important to note that two particularly exciting
candidate brown-dwarf events have been 
detected within  $6$~months of each other, both with $\tau_E < 16$~days. 
The first of these was detected 
serendipitously by an amateur astronomer,
A. Tago, 
searching for novae.  The Tago event was the first lensing event
discovered in a sparse field, through monitoring not
specifically designed to 
find evidence of lensing. The lensed source is an ordinary A0 star
located a kpc away, clearly indicating that the lens is nearby.
The event was of short duration, with the Einstein-diameter crossing
time estimated to be in the range $10-15$~days (Fukui et al.\, 2007)
and an estimated peak magnification  
of about $50.$ Gaudi et al.\, (2008) find that the most likely
explanation for this event is lensing by a nearby brown dwarf with
a proper motion greater than or approximately equal to
 $20$~mas~yr$^{-1}$.  
Because the source star was so bright, Fukui et al.\, (2007)
were able to verify one of the fundamental properties
of lensing predicted by Einstein's theory, that the
spectrum is unchanged by lensing. 
 In this case, we still know little about the lens, in spite of its apparent
proximity. During the course of the next decade, however, the lens and lensed
source should separate by a large enough angle that they can be resolved
by JWST, which is scheduled to be launched within the next 5 years.
The IR sensitivity of JWST, combined with its angular resolution will
allow it to detect the lens, if the lens is a brown dwarf. The angular 
separation achieved at the time of the observation will provide a
value of $\omega.$ The combination of $\omega$ and $\tau_E$ will yield
the value of $\theta_E.$ If the flux and spectrum of the brown dwarf 
allow the photometric parallax to be determined, the brown dwarf's mass
will be derived. Otherwise, a sequence of additional images will
determine the  geometric parallax, thereby allowing
us to measure the lens mass.

Six months after the Tago event, a second short-duration event that was
 most likely caused
by a brown dwarf was observed. 
This event had a peak magnification larger than $1000$. Both parallax
and finite-source-size effects were detected, allowing the mass 
($0.056 \pm 0.004\, M_\sun$) and distance to the lens 
($525 \pm 40$~pc) to be determined. The transverse velocity of the lens
is $113 \pm 21$~km~s$^{-1}$ (Gould et al.\, 2009). 

The discovery of these two brown dwarf events is remarkable because
each event had a high magnification, and therefore required 
a very small distance of closest approach. Such events are therefore rare. 
Each of the two events therefore represents
a large number of additional brown-dwarf events of short duration.
It would be difficult to use the detection of these
two events to formulate a realistic estimate of the total number of
short-duration brown-dwarf-lens events presently expected.
Nevertheless, these detections 
add plausibility to the estimates
we have made above,
 based on rate calculations and the combined OGLE and MOA
detection rates. 
  
Figure 1 demonstrates that 
brown-dwarf events with $\tau_E$
in the range of $8-16$~days can take place at 
distances greater than a hundred pc for velocities 
in excess of $\sim 50~$~km~s$^{-1}$.
Note in addition, that 
the total volume, hence the number of possible lenses 
and the rate of lensing by any
given population of lenses, increases  with distance  
from us. (See, e.g., Di\thinspace Stefano 2008a, 2008b for details.) 
Therefore
the largest number of brown-dwarf lenses generating
$8-16$-day events should have velocities $> 50~$~km~s$^{-1}$ and
be located at distances larger than $100$~pc.
This is consistent with the events observed to date. 

\subsubsection{An ideal planet-lens case} 

Consider a planetary system in which one planet serves as
a lens, producing a short-duration event with a measured value of $\tau_E.$
 Suppose that the star orbited by the planet lens
is detected. 
Suppose further, that a sequence of high-resolution measurements
allows the geometric parallax, proper motion, and Einstein
angle of the star to be measured (see, e.g., Di~Stefano 2009). 
The combination of $D_L$ and
$\theta_{E,\ast}$ produces a high-precision value of the 
gravitational mass, $M_\ast,$ of the star.
In general, the stellar mass is estimated based on spectral and flux 
information. A direct measurement of the gravitational mass allows
stellar models to be tested. 
In some cases, it may be possible to conduct subsequent
transit or radial-velocity studies to measure the gravitational mass
of the star in a second way, i.e., by studying the orbit
of the planet that served as a lens and/or the orbits of
other planets. Thus, for some stars orbited by planet lenses, we may
be able to compare the gravitational mass measured via lensing with the
gravitational mass measured via orbital dynamics.

Up to this point in our discussion, the planet has played
only a peripheral role: (1)~it produced a photometric event that alerted us to
the possibility of measuring astrometric lensing by the star, and (2)~it
alerted us to the presence of at least one planet orbiting the star, thereby
motivating subsequent transit and/or radial-velocity studies.
The lensing event can of course teach us a good deal about the planet.

First, if finite-source-size effects are detected, then $\theta_{E,planet}$ 
can be
directly measured. The distance to the planet is, to high precision, the same
as the distance to the central star. The combination
of $\theta_{E,planet}$ and $D_L$ measures the mass of the planet.
With the gravitational masses of both the planet and star measured,
the mass ratio $q$ can be computed. 

If, in addition, the projected orbital separation, $a,$ between the central
star and planet lens is less than roughly $3.5\, R_{E,\ast},$ 
or if the event ``repeats'', 
then
$q$ and $a$ can both be estimated from a fit to the planet-lens light curve.
The value of $q$ so measured can be checked for consistency with
the measured values of the planet's and star's gravitational masses.

The projected orbital separation, $a,$ measured from the light curve,
can be related to a combination of the true orbital separation at the
time of the event and the orbital inclination: $a=a_{true} cos(\theta).$
The value of $a_{true}$ is 
related to the orbital speed at the time of the event:
$v=30\, {\rm km/s} \times \sqrt{(M_\ast/M_\odot)\, ({\rm AU}/a_{true})}.$  

The proper motion of the planet can be measured from a combination of 
 $\theta_{E,planet}$ and $\tau_E$. With $D_L$ known, we can
estimate the projected value of the planet's orbital
velocity, $v\, cos(\theta),$  by  comparing the values
of $\omega_{planet}$ and $\omega_{\ast}.$
Combining the  
equations for $v$ and $a,$ yields a value for the inclination of the
orbit.

This example illustrates that for nearby lenses, orbital solutions can be
obtained, even for face-on orbits. Furthermore, there can be enough
information to provide independent checks on the values of some
physical parameters. 
Even 
planets in very wide orbits can be well studied with lensing, especially 
if there is a repeating event (Di\thinspace Stefano \& Scalzo 1999b).

\subsubsection{High-velocity Stars}

When high-velocity stellar-mass objects are nearby, their Einstein
angles are large enough to be measured during the event by 
measuring the centroid shift in the lensed source. 
In general, this leaves a degeneracy between $D_L$ and $M$.
The degeneracy can be resolved if the lens is detected.
Consider, for example, a halo dwarf star with $M=0.25\, M_\odot$ 
and $v=180$~km~s$^{-1}$.
If this lens is $34.8$~pc away from us, it can cause an event
with $\tau_E=5$~days. Assuming that $D_S>>D_L,$
The value of $\theta_E$ would be $7.6$~milliarcseconds, and the
astrometric shift induced by the lens could be
measured; the optimal case would be when the lensed source is a bright
blue star.  
Such a halo dwarf would itself be easily detected
after moving from the source position, and its proper motion
and parallax could be directly measured. Its gravitational mass
could therefore be measured to high precision.
This lens would travel across the sky with an angular speed 
of $\sim 1^{\prime\prime}$~yr$^{-1}$.  
With such a high angular speed. it is very likely that, if the background
field is dense, additional lensing events will occur over a time interval of
$\sim 10$~years. 
Because the presence of the lens is already known, it is easier
to identify future events with confidence, even if the 
angle of closest approach is larger. Once a second event is discovered,
the general direction of the lens motion is known and future
photometric and astrometric events are more easily predicted.
If a sequence of events is detected, the proper motion and parallax can
be computed by comparing the locations and times of the events.
This means that even for dark high-speed lenses, mass measurements can be 
made. When the lens is a neutron star, it may radiate x-rays as it cools and/or
accretes matter from the ISM; it may therefore be catalogued 
as a weak x-ray source.  
It is worth noting that repeating events have been observed (Skowron 
et al.\, 2009).

\section{Science, Strategies and Prospects}

\subsection{Science Goals} 

Every event of short duration is likely to be associated with
an interesting lens: a low-mass object,
or a high-velocity mass.
These events should therefore be high-priority targets for
intense monitoring and follow-up, even when the value of the peak  
magnification is modest.  
The choice of the range of values of $\tau_E$ on which to spend the greatest
resources 
must be governed by the science goals of the investigators.

\subsubsection{Planets} 
While any set of short events is likely to include planet-lens events, 
the set of the shortest events, those 
with  $\tau_E$ less than roughly $4-8$~days, is  
likely to consist primarily of planet-lens events.   
The returns for studying these events of very short duration 
are potentially large. 
Because the 
numbers of planets with $a > 1.5\, R_{E,\ast}$ is 
larger than the numbers in the narrow annulus within which
the so-called ``resonant'' events are generated,
the rate of planet discovery by 
lensing 
will increase when concerted efforts
to discover and study short events are made. 
We note further that 
the wide-orbit planets are almost certainly augmented by a 
significant number of free-floating planets. 

The potentially large contribution of 
wide-orbit planets to short and repeating lensing events 
was noted more than a decade ago (Di~Stefano \&
Scalzo 1999a, 1999b). At that time, however, the set up
of the monitoring and alert observing programs was not
well suited to the study of short-duration
and repeating events.  
A decade of advances in the observing programs
has put in place the ingredients needed to detect and identify
short-duration and repeating events.
Furthermore, the more sensitive photometry used today 
will play an important role in increasing the
detection efficiency 
Di~Stefano \& Scalzo 1999a, 1999b). 

In addition, as we note in the appendix, in the near
future, short events could begin to be routinely
identified in cases in which the planetary separation is small: 
$a<0.5\, R_{E,\ast}.$ 
These events present the exciting possibility
of using lensing to find nearby planets in the habitable zones 
of their central stars (Di~Stefano \& Night 2008).

In fact, as we have shown in the companion paper, choosing short-duration events
tends to select nearby lenses as well as lenses very close 
to the source star. The nearby planets that are identified through
their actions as lenses can be among the best-studied
and will become touchstones in the field of planetary studies.
  
\subsubsection{Brown Dwarfs and High-velocity Stars} 

Events of $8-20$~days 
are more likely to have been induced by brown dwarfs or fast moving
stars. 
The discovery of nearby neutron stars would, by itself,
be of great scientific value. 
The same is true of hypervelocity stars, runaway stars, and their remnants. 
Gravitational mass measurements of
neutron stars and other stellar remnants, as well as of
brown dwarfs, would advance our knowledge of fundamental science.
The feasibility of mass measurements has been demonstrated in 
several cases (In addition to Gould 2009, see
Alcock et al.\, 2001; 
Gould 2004;
Gould, Bennett,
\& Alves 2004; Drake, Cook, \& Keller 2004; Nguyen et al.\, 2004.)
The mass functions of brown dwarfs and stellar remnants
 have not yet been well established.
In addition, having a set of nearby objects would allow more detailed studies of
the relevant equations of state.

\subsection{Observing Programs}

Because monitoring teams are already identifying
short-duration events, identification 
of planets, brown dwarfs,
and high-velocity stars can start immediately.
This is demonstrated by Gould et al.\, (2009), although
for an event that had a duration on the longer end of
``short-duration'' events.  
Current programs could discover, every year, $45-69$
short events 
caused by brown dwarfs, $23-36$ short events
caused by wide-orbit planets,
$8-13$ short events caused by high-speed neutron stars,
and perhaps additional events caused by hypervelocity stars
or their remnants. 
Comprehensive model tests of a large fraction of these
events, including the planet events, will be possible.
Nearby planetary systems (12-18 per year) may be susceptible to
follow-up observations, including radial velocity or
transit studies. 

Pan-STARRS is about to begin taking data over a large region
using several sampling strategies (Chambers et al.\, 2004).  For example, 
regular monitoring will take place over 
roughly 70 square degrees in its Medium Deep
Survey. Additional wide-field coverage will eventually be 
provided by LSST, which will cover $\sim 20,000$ sq.\, degrees
every three to four nights.
While the cadence of these programs  is not ideal for 
the detection of short events, the efficiency for the 
detection of events caused by brown dwarfs, high-velocity
stars, and the most massive planets will be
high. Once identified by wide-field monitoring programs,
more frequent follow-up is required to fully characterize the light curve.  
For events of shorter duration, the efficiency of event
detection and identification will be lower.  
If, however, as part of each independent 
scan, different filters are used in a sequence over a time interval
during which the magnification is changing, even very
short-duration events could be
identified as targets for more frequent monitoring. 
The maximum value of the efficiency would be: 
$f\, \tau_E/\Delta t,$ where $f$ is a number in the range $0.1-1,$
$\Delta t$ is the time between distinct sequences of observations.
In LSST and portions of the Pan-STARRs fields, $\Delta t$ 
will be $3-6$ days.  
Even efficiencies of a percent are significant, because the 
total area monitored will be large enough that many lensing events are
expected, in front of both dense and sparse fields (Di~Stefano 2008b, Han 2008).
  
It is likely that, within five years, 
a program ideal for the 
detection of short-duration events will begin.  
Funding was recently approved for the {\it Korean
Microlensing Telescope Network (KMTNet)}, which
plans to use $3$ telescopes across the Southern
Hemisphere to provide monitoring with
10-minute cadence of approximately $16$~sq. degrees
(Han 2009).
As Figure 2 shows,
KMTNet can provide excellent monitoring for Earth-mass planets. 
It could
even discover dwarf planets with 
Einstein-diameter crossing times
of less than an hour, although in such cases, follow-up by other
observers would be required to obtain good sampling.
\subsection{Protocols} 

Below we summarize the steps that can be taken in the 
immediate future by monitoring teams already in action.  
At the end of the section we discuss work that could be conducted with
archival data. 
\begin{itemize}

\item The first step is to increase the effective cadence, to increase the
chances of identifying events of short duration.  
  Cooperation among teams, e.g., the MOA and OGLE teams
could effectively increase the cadence of monitoring
in selected fields.
\end{itemize}

The philosophy of microlensing monitoring 
has been based on the premise that events are generated by lenses
that are not known {\it a priori.}  For mesolenses, though, this is
not always the case. For example, when the lens producing a short-duration event 
happens to be  
a planet in orbit with a nearby star, 
there is a good chance that 
the star
has already been detected, especially when the region has been 
monitored over an interval of years.

\begin{itemize}
\item  To facilitate the identification of all events
in which radiation from a component of the lens system
can be detected, it is important for monitoring programs to
develop automatic links
to multiwavelength  
data bases that catalogue objects that might correspond to
the lens. Stellar remnants may have been 
detected at X-ray wavelengths, brown dwarfs in the infrared,
stars orbited by planets at optical and/or infrared. 
We note that, although this step is important for short events,
it is also potentially very useful even for longer events.
Whatever the event duration, this step will identify those
events caused by nearby lenses, facilitating
high-precision measurement of the lens mass. 
Ideally, in addition to the catalogues, 
the monitoring teams should have
quick access to the data taken at various wavelengths, 
since many 
nearby sources may not have been catalogued. 

\item When a short-duration event is caused by a planet orbiting
a star, and when the planet-star separation is less than
roughly $3.5\, R_{E,\ast},$ the presence of the star would 
already have caused a gradual increase in the magnification,
even before the short-duration event caused by the planet.
Monitoring teams can test for such a prior 
trend at the start of what appears to be
a short-duration event.

\item ``Repeating'' events in which a planet and its star
serve as independent lenses are expected in 
as many as $\sim 10\%$ of events in which a star with 
a planetary system serves as a lens. 
(See Di\thinspace Stefano \& Scalzo 1999b.)  
 The most likely case is the one in which the planet lens producing the
shorter event of the two events is the innermost ``wide'' planet.
 Roughly half the time,
the star should have produced the first event. Any short
duration event that follows a stellar-lens 
event has a good chance of having been produced
by a planet in orbit with the star. 
It would be worthwhile to subject the declining portion of
each stellar-lens event to monitoring capable of quickly 
identifying subsequent short events. For planets with masses
comparable to or larger than the that of the Earth,
monitoring $1-3$ times per day would be adequate to 
identify any additional perturbations that occur either as the
stellar-lens event declines or with $\tau_E$ of the 
return to baseline. An alert should be called immediately
when such deviations are detected.
It is important to note that systematic monitoring of this type
will either yield planet detections (the most likely outcome)
or will allow meaningful limits on the structure of 
planetary systems to be derived. 

\item High-resolution images can play an important role for a wide range
of lenses producing short-duration events. The possible uses of such
imaging are described in \S 3 and in the companion paper. In some cases, it
is important to obtain an image 
soon after the event is identified. In others, images taken
taken  years after the event may be useful. By using the light curve 
sampling and
archived data to identify the most promising models, observers can
identify those cases in which HST or JWST may provide
key information about the lens.    
\end{itemize}

\begin{itemize}
\item  
When
an ongoing event is identified as a possible short-duration event,
an alert should be called. 
Ground-based multiwavelength observations with good angular resolution, 
taken as the magnification increases, can
establish the amount of blending and  
determine if the
short duration of the event is an artifact caused by blending.
If so, monitoring for the reasons explored in this paper
is no longer necessary.
Events that are genuinely short 
should be subject to intensive monitoring by a
global network of telescopes. In some cases, the monitoring programs
may themselves sample the light curve frequently enough to allow
good model fits. 
Model fits to a well-sampled light curve 
can determine the values of $\tau_E$, $b$, and discover
whether finite-source-size effects, the existence of a companion,
and even parallax effects 
influence the light curve shape. The cadence of monitoring should be
greatest near peak, even if the magnification is not high. 
\end{itemize}

Finally, we note that it is not always  possible to follow each event as 
it unfolds. This is certainly the case for events that have already occurred.
Even future monitoring programs, with their improved detection efficiencies, 
may not identify or be able
to follow all short-duration events in real time.
In these cases, light curve fits to the short-duration event,
including possible long-term deviations from baseline that could
``frame'' the short-duration event, can be carried out.  In addition, step 2
above (studying multiwavelength catalogs and data that cover the region around
the event), can provide important information. In some cases, follow-up
HST  imaging can also be used to test models, e.g., allowing the 
lens mass to be determined in some cases.

\section{APPENDIX: Very Close Planets and the Habitable Zone} 
In the context of planetary systems, independent events of short
duration were originally predicted for planets in wide orbits
($a$ greater than roughly $1.5\, R_{E,\ast}$).
Recently, however, close orbits ($a<0.5\, R_{E,\ast}$) were studied, 
and
a second class of short-duration event 
considered (Di~Stefano \& Night 2008). 

For small $a$, there are two regions in the lens plane where
the effects of the planet are significant. The first  is close
to the central star. Events in which the source track passes close
to the star therefore exhibit short-lived deviations from the point-lens
form. As the value of $a$ decreases below $0.5\, R_{E,\ast},$
the region of perturbations caused by the planet becomes
smaller. Consequently the deviations in the light curve are
shorter-lived and are also more easily washed out by finite-source-size
effects. The second region is at a distance of roughly $1/a$ from the
star, along the binary axis. (See Di~Stefano \& Night 2008). This region
has a small caustic, which has little effect on most light
curves. There is a region around 
the caustic, however,
 in which the deviation from the low-magnification effects of the 
star can be as large as a few percent, lasting from hours to days.

KMTNet will be very sensitive to these deviations. But even today's
monitoring programs could find them. To do so would be important,
because for nearby M~dwarfs, orbital separations of
$0.2-0.5\, R_{E,\ast}$ put the planet in the zone of habitability. 

The teams could call alerts on such events with a high level
of confidence because of the presence of the star along the line
of sight and also because a low-magnification effects associated
with the star, frame the short-duration deviation caused by
the planet. 
This is in exact analogy to the situation in which
a short-duration event is caused by a planet with
$1.5\, R_{E,\ast} < a < 3.5\, R_{E,\ast}.$  

There are several new features of lensing in this close-planet
regime. 

\noindent {\bf (1)} The deviation that signals the
presence of the planet is not generally well approximated
by a point-lens model.

\noindent {\bf (2)} The deviation that signals the presence of the planet
occurs when the source is far from the central star ($\sim 1/a$).

\noindent {\bf (3)} Because the planet is close to the star,
orbital motion can be significant during the event.
This means that the regions of deviation are likely to swing into the
path of the source star. We have found that the rate of these
potentially important events can be comparable to the rate of
stellar-lens events
exhibiting $A>1.34$.

We therefore expect that, ultimately, events in which we can find
evidence of a planet orbiting close to the central star,
potentially in the habitable zone, could be an important addition to
short-duration events.

\bigskip
\bigskip
\noindent {\bf Acknowledgments:} This work benefited from discussions
with  
Charles Alcock, Erin Arai, Mary Davies, 
Nitya Kallivayalil, M.J. Lehner, Christopher Night, Brandon Patel, 
Frank Primini, Pavlos Protopapas, Rohini Shivamoggi,  
Kailash Sahu, and Takahiro Sumi. 
I would also like to thank the anonymous referee
for useful comments.   
The research was conducted under the aegis of NSF grants
AST-0708924 and AST-0908878, and 
a grant from the Smithsonian endowment. I would like to thank the
Aspen Center for Physics for its hospitality during the early
phases of this work.  

\bigskip
 
\centerline{\sl References} 

\noindent
Alcock, C., et al.\
2001, \nat, 414, 617










\noindent
Bond, I.~A., et al.\ 2001, MNRAS, 327, 868
https://it019909.massey.ac.nz/moa/



\noindent
Chambers, 
K.~C., \& Pan-STARRS 2004, American Astronomical Society Meeting Abstracts, 
205,  


\noindent
Di Stefano, R.\, 2009, arXiv:0912.1611  

\noindent
Di Stefano, R.\, 2008a, ApJ, 684, 46

\noindent
Di Stefano, R.\, 2008b, ApJ, 684, 59

\noindent
Di Stefano, R., \& Esin, A.~A.\ 1995, \apjl, 448, L1


\noindent
Di Stefano, R., \& Scalzo, R.~A.\ 1999b, \apj, 512, 579 

\noindent
Di Stefano, R., \& Scalzo, R.~A.\ 1999a, \apj, 512, 564 

\noindent
Di Stefano, R. \& Night, C.\, 2008, ArXiv e-prints, 0801.1510



\noindent

\noindent 
Drake, A.~J., Cook, 
K.~H., \& Keller, S.~C.\ 2004, ApJL, 607, L29 

\noindent
Fukui, A., et al.\ 2007, \apj, 670, 423

\noindent
Gaudi, B.~S., et al.\ 
2008, \apj, 677, 1268



\noindent
Gould, A., et al.\ 2009, 
arXiv:0904.0249

\noindent
Gould, A., Bennett, 
D.~P., \& Alves, D.~R.\ 2004, ApJ, 614, 404 

\noindent Gould, A.\ 2004, ApJ, 606, 319 


\noindent 
Griest, K., \& Safizadeh, N.\ 1998, \apj, 500, 37


\noindent
Near-Field
Han, C.\ 2008, \apj, 681, 806

\noindent
Han, C.\ 2009, private communication  

\noindent

\noindent
Skowron, J., 
Wyrzykowski, {\L}., Mao, S., \& Jaroszy{\'n}ski, M.\ 2009, \mnras, 393, 999



\noindent
Udalski, A. \, 2003, Acta Astron., 53, 291


\begin{figure*}
\begin{center}
\psfig{file=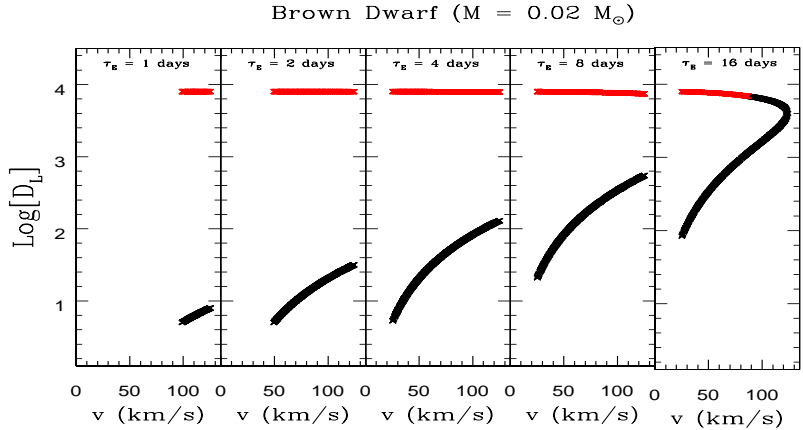,
height=4.5in,width=4.5in,angle=-0.0}
\vspace{-2.1 true in}
\caption{
Logarithm of $D_L$ vs $v$ for a lens with fixed mass ($0.02\, M_\odot$)
in the brown-dwarf regime. 
The field of background sources was placed at a distance $D_S$ of $8$~kpc. 
Each lens
was given a randomly generated value of the transverse velocity, $v,$
in the range $25-125$~km~s$^{-1}$. We assumed that the Einstein-diameter
crossing time was $1$~day in the leftmost panel, 
increasing by a factor of two in each panel to the right. 
 For each lens, there are two
possible values of $D_L,$ $D_L^+$ and $D_L^-$. 
We computed both and show the results.
If, for $D_L^+,$
 the value of $b\, \theta_E \, D_S$ (with $b=0.1$) was smaller than
$10\, R_\odot,$ we assumed that finite-source-size effects could be detected and plotted the point in red.}
\end{center}
\end{figure*}

\begin{figure*}
\begin{center}
\psfig{file=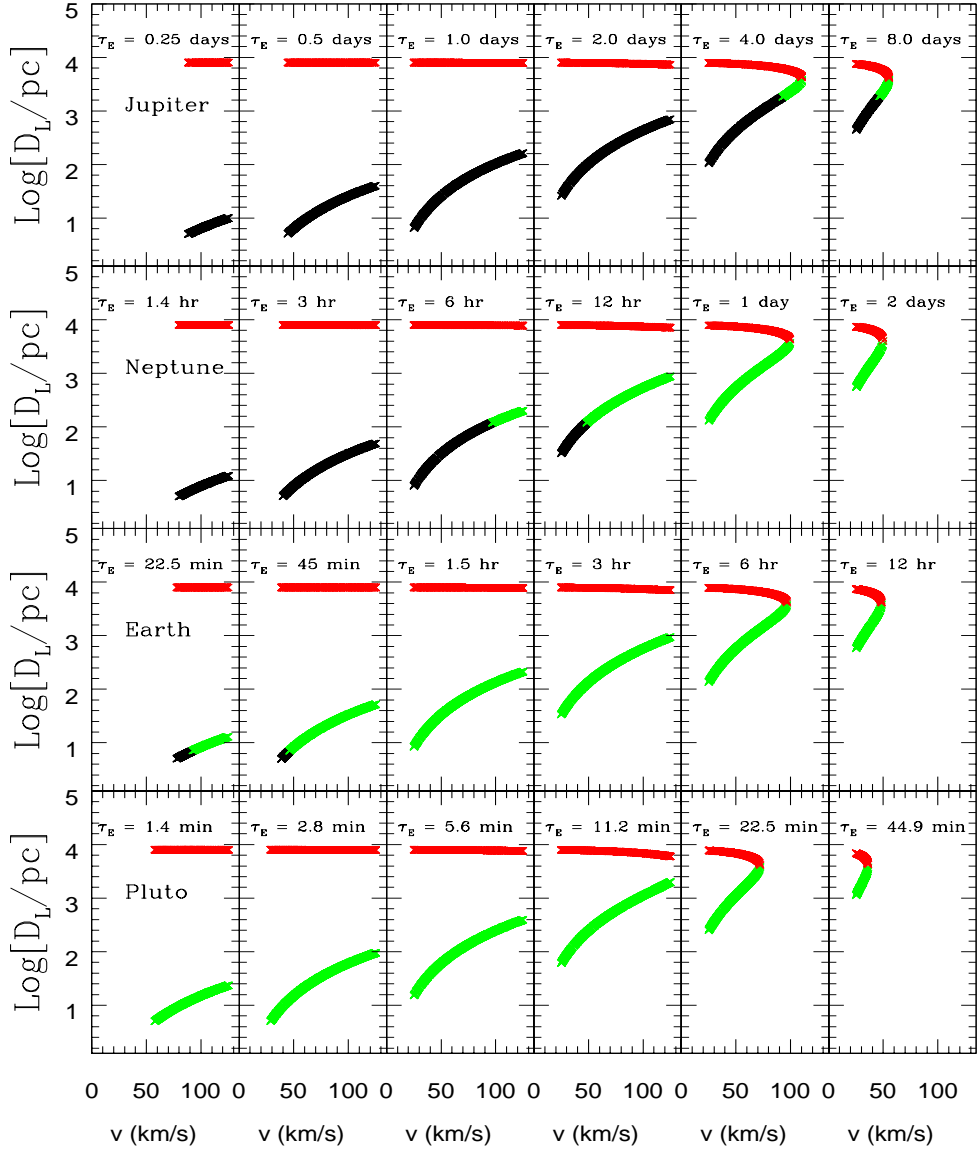,
height=6.5in,width=5.5in,angle=-0.0}
\vspace{1.0 true in}
\caption{Each row consists of a set of $6$ panels in which the  
logarithm of $D_L$ vs $v$ is plotted for a lens with fixed mass.
The mass chosen is that of the planet which labels the row. 
For each planet the range of interesting time scales is shown.
Note how this range is different for planets of different mass.
If, for $D_L^+, (D_L^-)$ 
 the value of $b\, \theta_E \, D_S$ (with $b=0.1$) was smaller than
$10\, R_\odot,$ we assumed that finite-source-size effects could be detected and plotted the point in red (green).
See the caption of Figure ~1 for additional details. 
}
\end{center}
\end{figure*}

\begin{figure*}
\begin{center}
\psfig{file=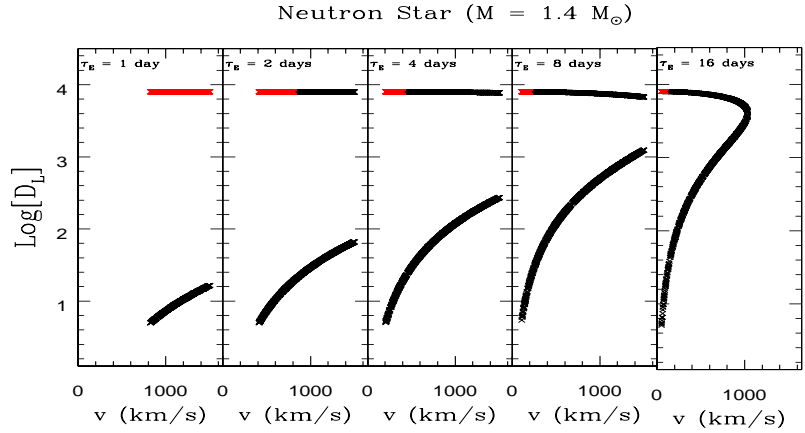,
height=4.5in,width=4.5in,angle=-0.0}
\vspace{-2.1 true in}
\caption{
Logarithm of $D_L$ vs $v$ for a lens with fixed mass ($1.4\, M_\odot$). 
The Einstein diameter crossing time was $1.4$~hours 
in the leftmost panel, 
increasing by a factor of two in each panel to the right. 
See the caption of Figure ~1 for details.} 
\end{center}
\end{figure*}

\end{document}